# Critical currents and vortex dynamics in superconducting MgB$_2$

Y. Bugoslavsky, G.K. Perkins, X.Qi, L.F. Cohen and A.D. Caplin

*Centre for High Temperature Superconductivity, Blackett Laboratory, Imperial College, London SW7 2BZ, UK*

The recently-discovered[1] MgB$_2$ superconductor has a transition temperature $T_c$ approaching 40K, placing it intermediate between the families of low and high temperature superconductors (LTS and HTS). In practical applications, super-conductors are permeated by quantised magnetic flux vortices, and when a current flows there is dissipation unless the vortices are "pinned" in some way, and so inhibited from moving under the influence of the Lorentz force. This vortex motion sets the limiting critical current density $J_c$ in the superconductor. Vortex behaviour has proved to be more complicated in the HTS than in LTS materials. While this has stimulated extensive theoretical and experimental research,[2] it has impeded applications. Clearly it is important to explore vortex behaviour in MgB$_2$; here we report on $J_c$, and also on the creep rate $S$, which is a measure of how fast the "persistent" currents decay. Our results show that naturally-occurring grain boundaries are highly transparent to supercurrent, and suggest that the steep decline in $J_c$ with increasing magnetic field $H$ reflects a weakening of the vortex pinning energy, possibly because this compound forms naturally with a high degree of crystalline perfection.

A negative feature of the HTS phases is that the grain boundaries impede the flow of supercurrent. This problem is particularly severe in the YBa$_2$Cu$_3$O$_7$ phase, which intrinsically has the highest $J_c$ of the HTS oxides, but behaves poorly in bulk polycrystalline form, with $J_C$ depressed by several orders of magnitude at even modest-angled grain boundaries.[3] Thus measurements on polycrystalline $MgB_2$ - the only form available so far - have to be analysed carefully.

We extracted from $MgB_2$ powder (Alfa Aesar Co., 98% purity) several fragments of above 100 micron size; they are somewhat irregular in shape, and XRD indicates no obvious crystalline texture. The superconducting transition (Fig. 1(a)) of a single particle is sharp with a $T_c$ of 37 K, and (Fig. 1(b)) they have well-defined tunnelling characteristics[4] with a gap $2\Delta$ of ~5 $k_B T_c$ (for a BCS superconductor this factor is 3.5, for the HTS phases $Bi_2Sr_2CaCu_2O_8$ and $YBa_2Cu_3O_7$, it is 10 and 5.5 respectively[5,6]). The key data are derived from magnetisation loops that have excellent signal-to-noise (Fig. 1(c)). Before proceeding further, we applied a number of cross-checks that have been developed in the context of HTS studies.[7] The initial slopes *(dm/dH)$_{init}$* of the *m(H)* loops are a measure of the volume from which screening currents exclude magnetic flux; likewise, when the field sweep direction is reversed at an extremum of the loop, the "reverse leg" slope *(dm/dH)$_{rl}$* measures the volume over which the screening currents attempt to maintain constant flux. If the sample is a single well-connected entity, these slopes should be equal to the actual sample volume (apart from an enhancement associated with classical demagnetising effects). Our data yield measured slopes of 7.4±0.4 x $10^{-12} m^{-3}$, independent of field and temperature. In an SEM, this sample appeared to have roughly ellipsoidal shape, with cross-section axes of 450 and 250 microns, and between 100 and 150 microns in the third direction; its physical volume $\Omega$ is therefore between 5 and 7 x $10^{-12} m^{-3}$. The screened volume might be up to a factor of 1.5 greater than this because of demagnetising effects, so within the uncertainties involved, there is excellent agreement with the measured slope. Thus, the straightforward conclusion is that screening currents flow in a well-connected fashion around this small polycrystalline fragment, and that over the field and temperature range studied, the grain boundaries are transparent to supercurrent, just as in LTS materials. In contrast, a polycrystalline sample of $YBa_2Cu_3O_7$ (unless very well-textured crystallographically) would behave as an assembly of disconnected grains at all but the lowest applied fields.

In order to convert the irreversible magnetic moment to $J_C$ we treat the fragment as a disc of radius $a$ of 300 microns and thickness $t$ of 100 microns, i.e. a volume $\Omega$ of 7 $10^{-12}$m$^{-3}$, at the upper range of our estimate, so as to yield conservative values for $J_C$. The standard Bean model gives: $J_C = 3\Delta m / 2\Omega a$ (in SI units, but in keeping with standard practice in the field, we report $J_C$ in A cm$^{-2}$ rather than A m$^{-2}$), where $\Delta m$ is the full-width of the $m(H)$ loop. The $J_C$'s (Fig. 2) are high; as a benchmark, at 20 K and 1 T, it is 1.0 x 10$^5$ A cm$^{-2}$. However, $J_C$ decreases quite rapidly with applied field, and an "irreversibility field" $H_{irr}$ (Fig. 2, inset) can be defined above which $J_C$ drops below the noise level, in our case ~10$^3$ A cm$^{-2}$. Because we have established that the sample is well-connected, the $J_C$ shown in Fig. 2 is a reliable measure of the *intra*granular critical current density in MgB$_2$.

There have already been several magnetic studies of pressed and sintered pellets of MgB$_2$.[8,9,10] All the irreversibility fields are similar to the values shown in Fig. 2, but in two cases[8,9] the $J_C$'s are more than an order of magnitude less than ours. Takano et al.,[10] who sintered their material at high temperatures and pressures, report a rather larger $J_C$, ~4 x 10$^4$ A cm$^{-2}$ at 20 K and 1 T,[*] but still a factor of 2 or 3 less than we measure. Overall, these results can be understood straightforwardly: in sintered pellets of MgB$_2$ the intergrain contact is less perfect than in our polycrystalline as-grown fragments, to a degree dependent upon processing; the reduced cross-sectional contact area limits the magnitude of the macroscopic $J_C$, but not its field and temperature dependences.

---

[*] In addition, they measured MgB$_2$ powder and converted magnetisation to $J_C$ by assuming a uniform 1 micron particle diameter, obtaining $J_C$'s an order of magnitude larger still; however, this result should be treated with caution, because with a particle size distribution, the magnetic moment is heavily weighted by the larger grains, leading to an over-estimate of $J_C$.

The other key aspect of the vortex physics is that because of thermal activation, the vortices are never totally immobile, and the screening current and associated magnetic moment decay with time in a quasi-logarithmic fashion. The relevant parameter is the creep rate $S$, defined as $-\partial \ln m / \partial \ln t$. Equivalently,[11] the amplitude of the $m(H)$ loops depends on the field ramp rate $\dot{H}$ (which is proportional to the electric field $E$ induced in the sample), and to a good approximation $S = \partial \ln m / \partial \ln \dot{H}$. As shown in Fig. 3, $S$ is small at low fields, but increases quasi-exponentially as the field approaches $H_{irr}$. The creep rate is related directly to the current-voltage characteristic of the material, with $E \propto J^{1/S}$; thus in $MgB_2$ the *E-J* characteristics at low fields are very steep, which is desirable for applications. At fields above $H_{irr}$, the vortex response becomes linear, $E \propto J$, or equivalently $S=1$; consequently, it is to be expected that at fields close to $H_{irr}$, $S$ should approach unity.

The two quantities $J_C$ and $S$, measured over a range of field and temperature, together provide the essential information for an understanding of the vortex behaviour. Rather than $J_C$ itself, it is better to consider the dimensionless derivative $\partial \ln J_c / \partial \ln H$, i.e. the slope of the curves of Fig. 2, which we denote as χ. A plot of χ (at a fixed temperature) against $S$ with $H$ as the implicit parameter has a direct physical interpretation:[12] In the HTS materials, these plots are linear, with slopes almost independent of $T$ (except close to $T_c$). This linearity indicates simple power law dependences on $H$ of two fundamental quantities, the current density $J_0$ which would be attained if there were no thermal activation of the vortices, and an energy $U_0$ which measures the vortex pinning as modified by vortex-vortex interactions. In $MgB_2$, these plots (Fig. 4) are again close to linear, and at 35 K (i.e. close to $T_c$) have almost the same slope as that seen in $YBa_2Cu_3O_7$. However, as the temperature drops, they become considerably steeper, which indicates that in $MgB_2$ at temperatures not too close to $T_c$, $U_0$ decreases very rapidly with increasing field. Unfortunately, this kind of detailed information is not available for conventional LTS superconductors, so that no comparison can be made.

MgB$_2$ is certainly a promising material in some respects, $J_C$ is high and grain boundaries appear benign, but the irreversibility field is rather low, only about half of $H_{c2}$.[8,9] The sharp rise in the creep rate $S$ as the field approaches $H_{irr}$ and the steep χ-$S$ plots suggest that the vortex pinning interactions weaken catastrophically with increasing field. Perhaps the high degree of crystalline perfection of this compound, as indicated by the very low residual resistivity[13] and by the reproducibilty of $T_c$ and $H_{irr}$ between samples from different sources, is the cause. For MgB$_2$ to become more useful for applications, it will be essential to find ways to enhance the vortex pinning energy.

1    Nagamatsu, J., Nakagawa, N., Muranaka, T., Zenitani, Y. & Akimitsu, J. *Nature* (2001).(in press)

2    Cohen, L.F. & Jensen, H.J. *Rep.Prog.Phys.* **60**, 1581-1672 (1997)

3    Mannhart, J., Chaudhari, P., Dimos, D., Tsuei, C.C. & McGuire, T.R. *Phys.Rev.Lett.* **61**, 2476-2479 (1988).

4    Y.Bugoslavsky et al. (in preparation)

5    Miyakawa N., Zasadzinski J.F., Ozyuzer L., Guptasarma P., Hinks D.G., Kendziora C., Gray K.E. *Phys. Rev. Lett.* **83**: 1018-1021 (1999)

6    Cucolo A.M., DiLeo R., Nigro A., Romano P., Bobba F., Bacca E., Prieto P. *Phys. Rev. Lett.* **76**   1920-1923 (1996)

7    Caplin, A.D., Cohen, L.F., Perkins, G.K. & Zhukov, A.A. *Supercond.Sci.Technol.* **7**, 412-422 (1994).

8    Larbalestier, D.C., Rikel, M., Cooley, L.D., et al.  *Nature* (2001).(in press)

9    Finnemore, D.K., Ostenson, J.E., Bud'ko, S.L., Lapertot, G. & Canfield, P.C.  *cond-mat* **/0102114**, (2001).

10   Takano, Y., Takeya, H., Fujii, H., et al.  *cond-mat* **/0102167**, (2001).

11   Pust, L., Jirsa, M. & Durcok, S.  *J.Low Temp.Phys.* **78**, 179-186 (1990).

12   Perkins, G.K. & Caplin, A.D.  *Phys.Rev.B* **54**, 12551-12556 (1996).

13   Canfield, P.C., Finnemore, D.K., Bud'ko, S.L., Lapertot, G., Cunningham, C.E. & Petrovic, C.  *cond-mat* /0102289 (2001).

**Acknowledgments**
We thank Dr. D. Cardwell for the supply of $MgB_2$, and Dr. A.J.P. White for his crystallographic advice. This research has been supported by the UK Engineering & Physical Sciences Research Council.

Correspondence should be addressed to Y.B. (e-mail: y.bugoslav@ic.ac.uk).

Figure 1. Superconducting response of a single MgB$_2$ fragment: (a) Magnetic moment as a function of temperature in a field of 10 mT (previously cooled in zero field), showing a sharp superconducting transition at 37 K. (b) Typical tunnelling characteristic on a similar fragment; the parabola is an estimate of the background conductivity. (c) Magnetisation loop (vibrating sample magnetometer, Oxford Instruments Model TVSM5) of the fragment shown in (a) at 20 K; data taken at a ramp rate of 13.3 mT s$^{-1}$; the broken lines show the slopes that provide an estimate of the sample volume.

Figure 2 Critical current density $J_C$ of the MgB$_2$ fragment as a function of magnetic field. The values shown are a lower bound, because of uncertainties in sample dimensions and shape, and could be up to about 50% larger. Inset: The irreversibility field $H_{irr}$ as a function of temperature. Above $H_{irr}$, irreversible (i.e. screening) currents cannot be sustained; a criterion of $J_C=10^3$ A cm$^{-2}$ is used here to define $H_{irr}$.

Figure 3 Creep rate $S$ as a function of field. These data are derived from the differences of irreversible magnetic moment between loops taken at ramp rates of 13 and 3 mT s$^{-1}$. Because the differences are small, the signal has been increased by using a sample of 10 MgB$_2$ fragments. The oscillations in the curves, particularly those at 25 K, are artefacts arising from small fluctuations in sample temperature, whose effects are amplified by the differencing of the data used to extract $S$. In the region below the broken line, the self-field (the field generated by the screening currents themselves) is comparable to or larger than the applied field, and so complicate the interpretation.

Figure 4 Vortex behaviour in MgB$_2$ compared with HTS materials, summarised by a plot of $\chi$, the dimensionless slope of the ln$J_c$ – ln$H$ curves of Fig. 2, against the creep rate $S$, with $H$ as implicit parameter at the temperatures indicated; the oscillations are instrumental artefacts, as described in the legend to Figure 3. The broken line represents the behaviour of the HTS superconductor $YBa_2Cu_3O_7$ at temperatures not too close to $T_c$.[12]

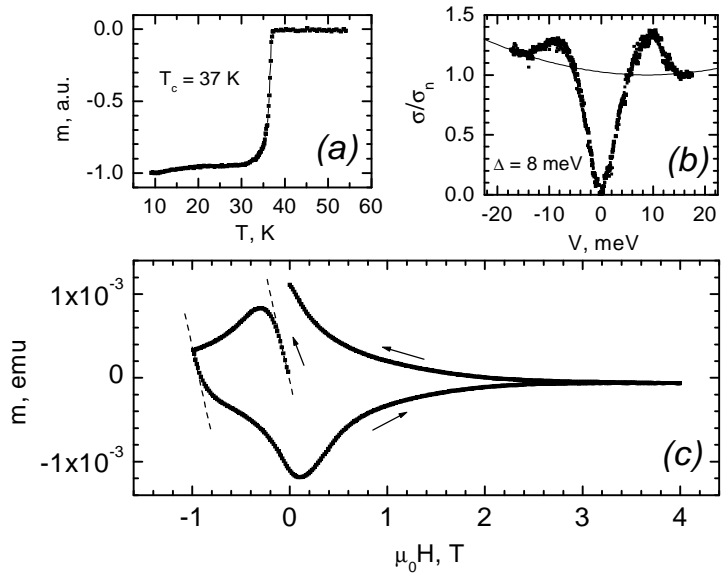

FIGURE 1

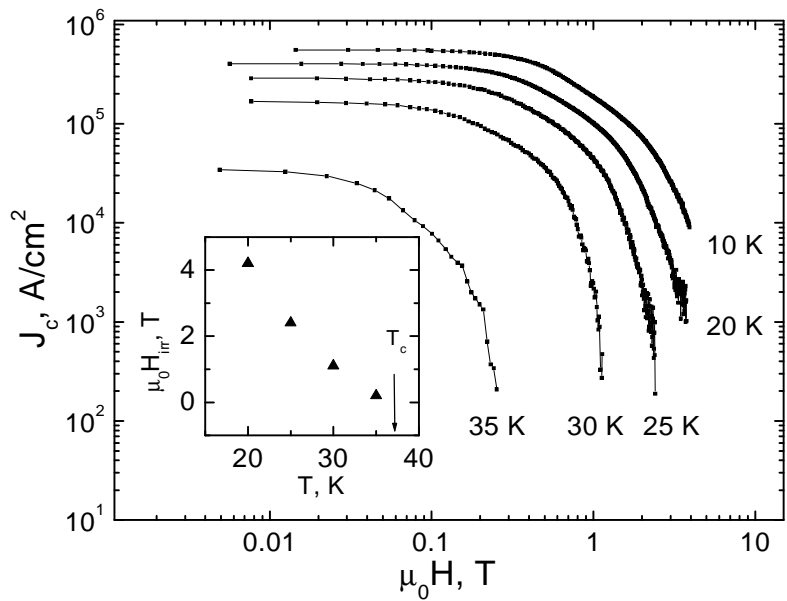

FIGURE 2

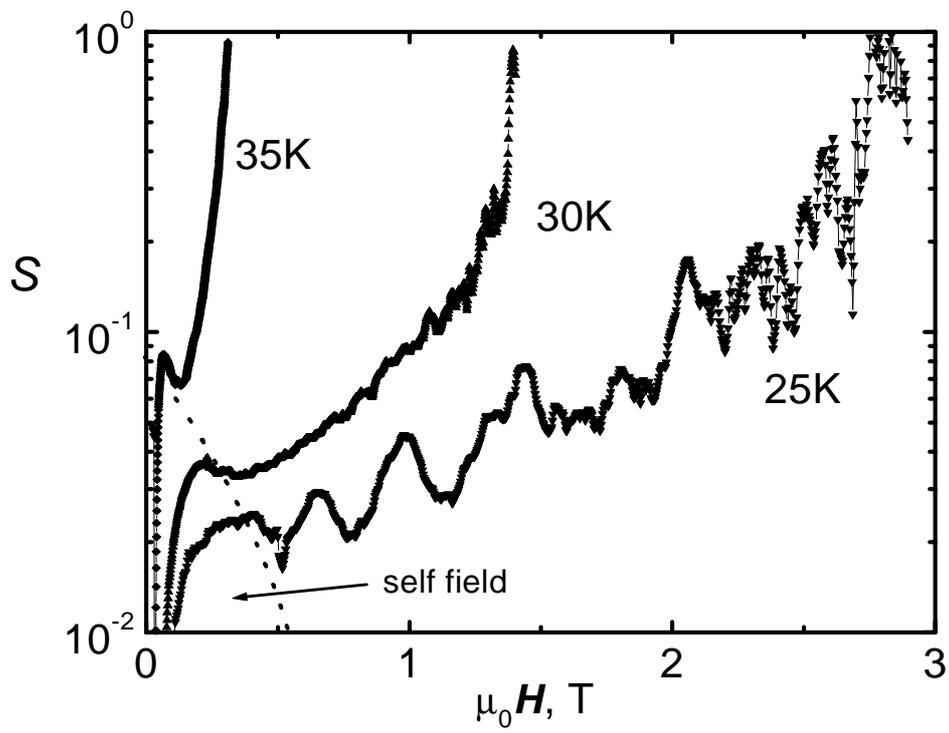

FIGURE 3

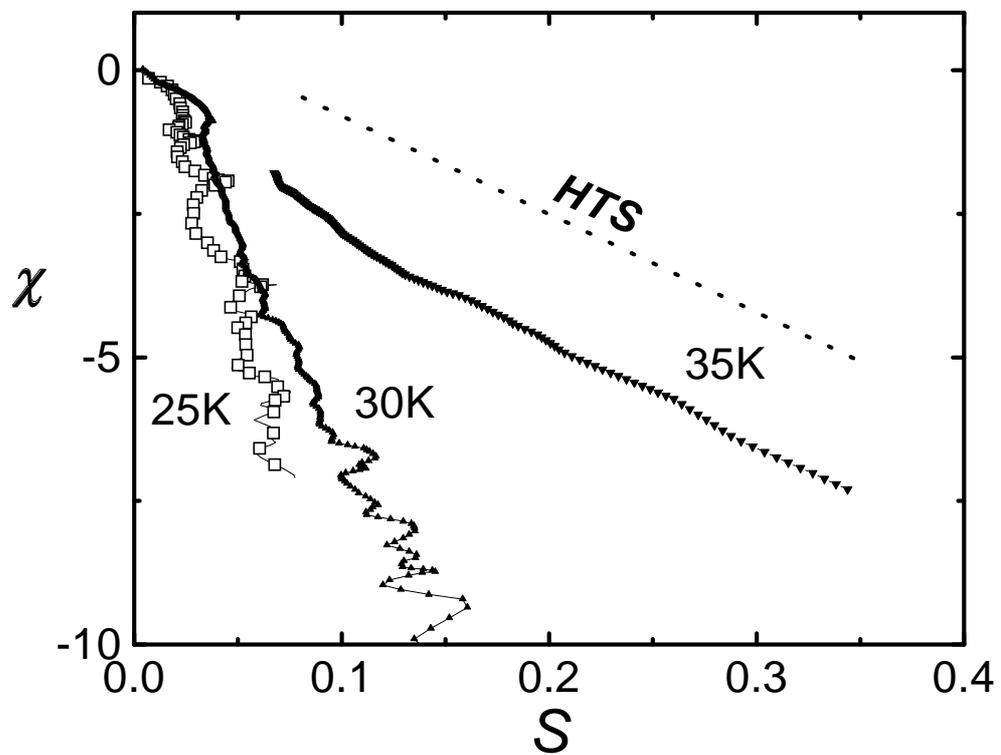

FIGURE 4